\documentclass[superscriptaddress,prb,preprintnumbers,nobibnotes,aps,preprint]{revtex4-1}

\pdfoutput=1

\usepackage{amsmath,amssymb}
\usepackage{float}
\usepackage{bm}
\usepackage{graphicx}

\usepackage{epstopdf}
\usepackage{color}
\usepackage{verbatim}
\usepackage{multirow}
\usepackage{textcomp}
\usepackage{dcolumn}
\newcolumntype{d}[1]{D{.}{.}{-1}}
\usepackage{natbib}
\usepackage{notes2bib}
\usepackage[hyperfootnotes=false]{hyperref}
\usepackage[hyperfootnotes=false]{hyperref}
  \hypersetup{colorlinks,
              citebordercolor=.1 .1 1,
              linkbordercolor= .1 .1 1,
              urlbordercolor=.1 .1 1, }

\begin{document}

\title{Revisiting the phase diagram of LaFe$_{1-x}$Co$_x$AsO on single crystals by thermodynamic methods}

\author{F.~Scaravaggi}
\email{f.scaravaggi@ifw-dresden.de}
\affiliation{Institute for Solid State Research, Leibniz IFW Dresden, 01069 Dresden, Germany}
\affiliation{Institute of Solid State and Materials Physics and W\"urzburg-Dresden Cluster of Excellence ct.qmat, TU Dresden, D-01062 Dresden, Germany}
\author{S.~Sauerland}
\affiliation{Kirchhoff Institut f\"ur Physik, Heidelberg University, 69120 Heidelberg, Germany}
\author{L.~Wang}
\affiliation{Kirchhoff Institut f\"ur Physik, Heidelberg University, 69120 Heidelberg, Germany}
\author{R.~Kappenberger}
\affiliation{Institute for Solid State Research, Leibniz IFW Dresden, 01069 Dresden, Germany}
\affiliation{Institute of Solid State and Materials Physics and W\"urzburg-Dresden Cluster of Excellence ct.qmat, TU Dresden, D-01062 Dresden, Germany}
\author{P.~Lepucki}
\affiliation{Institute for Solid State Research, Leibniz IFW Dresden, 01069 Dresden, Germany}
\author{A.~P.~Dioguardi}
\affiliation{Institute for Solid State Research, Leibniz IFW Dresden, 01069 Dresden, Germany}
\author{X.~Hong\footnote{Present address: Fakult\"at f\"ur Mathematik und Naturwissenschaften, Bergische Universit\"at Wuppertal, 42097 Wuppertal, Germany}}
\affiliation{Institute for Solid State Research, Leibniz IFW Dresden, 01069 Dresden, Germany}
\author{F.~Caglieris\footnote{Present address: CNR-SPIN, Corso Perrone 24, 16152 Genova, Italy}}
\affiliation{Institute for Solid State Research, Leibniz IFW Dresden, 01069 Dresden, Germany}
\author{C.~Wuttke}
\affiliation{Institute for Solid State Research, Leibniz IFW Dresden, 01069 Dresden, Germany}
\author{C.~Hess\footnote{Present address: Fakult\"at f\"ur Mathematik und Naturwissenschaften, Bergische Universit\"at Wuppertal, 42097 Wuppertal, Germany}}
\affiliation{Institute for Solid State Research, Leibniz IFW Dresden, 01069 Dresden, Germany}
\affiliation{Center for Transport and Devices, TU Dresden, 01069 Dresden, Germany}
\author{H.-J.~Grafe}
\affiliation{Institute for Solid State Research, Leibniz IFW Dresden, 01069 Dresden, Germany}
\author{S. Aswartham}
\affiliation{Institute for Solid State Research, Leibniz IFW Dresden, 01069 Dresden, Germany}
\author{S.~Wurmehl}
\affiliation{Institute for Solid State Research, Leibniz IFW Dresden, 01069 Dresden, Germany}
\affiliation{Institute of Solid State and Materials Physics and W\"urzburg-Dresden Cluster of Excellence ct.qmat, TU Dresden, D-01062 Dresden, Germany}
\author{R.~ Klingeler}
\affiliation{Kirchhoff Institut f\"ur Physik, Heidelberg University, 69120 Heidelberg, Germany}
\affiliation{Center for Advanced Materials, Heidelberg University, INF 225, D-69120 Heidelberg, Germany}
\author{A.~ U.~ B.~ Wolter}
\email{a.wolter@ifw-dresden.de}
\affiliation{Institute for Solid State Research, Leibniz IFW Dresden, 01069 Dresden, Germany}
\author{B.~B\"uchner}
\affiliation{Institute for Solid State Research, Leibniz IFW Dresden, 01069 Dresden, Germany}
\affiliation{Institute of Solid State and Materials Physics and W\"urzburg-Dresden Cluster of Excellence ct.qmat, TU Dresden, D-01062 Dresden, Germany}

\date{\today}

\begin{abstract}

In this work we revisit the phase diagram of Co-doped LaFeAsO using single crystals and thermodynamic methods. From magnetic susceptibility studies we track the doping evolution of the antiferromagnetic phase, revealing a continuous suppression of $T_\mathrm{N}$ up to 5\% Co doping. In order to study the evolution of the so-called nematic phase, the temperature dependence of the lengths changes along the $a$ and $b$ orthorhombic directions, $\Delta L/L_0$, was determined by high-resolution capacitance dilatometry. The results clearly show a gradual reduction of the orthorhombic distortion $\delta$ and of $T_\mathrm{S}$ with increasing Co content up to 4.5\%, while it is completely suppressed for 7.5\% Co. Bulk superconductivity was found in a small doping region around 6\% Co content, while both $T_\mathrm{c}$ and the superconducting volume fraction rapidly drop in the neighbouring doping regime. Ultimately, no microscopic coexistence between the superconducting and magnetic phases can be assessed within our resolution limit, in sharp contrast with other iron-pnictide families, e.g., electron- and hole-doped BaFe$_2$As$_2$.

\end{abstract}

\maketitle

\section{Introduction}

Since the discovery of high-temperature superconductivity in F-doped LaFeAsO by Kamihara \textit{et al.}\cite{KWH2008}, iron-based superconductors (IBS) were extensively studied and new families were soon discovered and successfully synthesized. The different compounds are classified by their stoichiometry, such that the most prominent families are denoted as 1111 (e.g., LaFeAsO\cite{KWH2008}), 11 (e.g., FeSe\cite{FeSe_SC_Hsu_2008}), 122 (e.g., BaFe$_2$As$_2$\cite{Ba122_review_Canfield_2010}) and 111 (e.g., LiFeAs\cite{LiFeAs_SC_Tapp_2008,LiFeAs_SC_Borisenko_2010}). The phase diagram of electron-doped 1111 systems has been extensively studied for polycrystalline samples over the past decade.\cite{FePn_Review_Stewart_2011,FLa1111_Nomura_2008,FLa1111_Luetkens,Hess_FLa1111_PC,Maeter_CeSr1111_arXiV_2012,Qureshi_neutron_2010} This family of compounds, $RE$FeAsO ($RE$= La, Ce, Sm), different from other iron pnictides, has well separated structural ($T_\mathrm{S}$) and magnetic ($T_\mathrm{N}$) transitions and shows a broad region of electronic nematic order below $\sim$~160~K. Bulk superconductivity can be induced by a small amount of electron doping and this family constitute the highest achieved superconducting transition temperatures ($T_\mathrm{c}$) to date among IBS.\cite{KWH2008,RE111_record_Tc_Ren_2008,Sm1111_Chen_2008,FCe1111_Chen_2008,FNd1111_Ren_2008}
According to several theoretical and experimental works, structural, magnetic and orbital degrees of freedom are closely intertwined and produce an overall breaking of the C$_4$ symmetry in the FeAs planes below $T_\mathrm{S}$, resulting in anisotropic behavior of several physical quantities within the $ab$ plane.\cite{Fradkin_nematic_fluids_review_2010,Fernandes_Nature_2014,Fernandes_PRB_preemptive_2012,Fernandes_Schmalian_2012,Chubukov_PRB_2015,Kang_FeSe_PRB_2018}
LaFeAsO proves to be particularly interesting for the study of such phenomena, because it is not influenced by an additional magnetic sublattice forming in the $RE$O layers coming from 4f electrons, compared to e.g., SmFeAsO. Previous studies on La1111 polycrystals demonstrated a strong suppression of the nematic phase with increasing F doping, showing a first-order-like transition as a function of electron doping (at $\sim$~5\%), and a substantial separation of the antiferromagnetic spin density wave (SDW) and the superconducting (SC) phase.\cite{Lang_NQR_RE1111_2016,Hajo_NMR_FLa1111_pc,FLa1111_mag_klingeler,R1111_TE_klingeler,La1111_SDW_klauss,McGuire_NJPhys_2009,Shiroka_LROtoSRO_Ce1111_2011}
This picture appears to be confirmed in the case of Co-doped LaFeAsO polycrystals, where the SDW/nematic transitions are only observable above $\sim$~100~K and disappear already at very low doping levels ($<$~2.5~\%). Superconductivity only develops above this threshold, suggesting competition between these phases.\cite{CoLa1111_PC_Sefat_2008,CoLa1111_PC_2009}

In this context, capacitance dilatometry proved to be a very sensitive tool to probe not only the small orthorhombic distortion, considered as the hallmark of the nematic transition, but also magnetic and superconducting transitions within the IBS families.\cite{BaNa122_Wang_2016,122P_Boehmer_2012,SrNa122_Wang_2019,BaK122_Boehmer_Nat_comm_2015,FeSe_review_Boehmer_2017,FeSe_orth_p_Kothapalli_2016,Boehmer_TD_NMR_FeSe_2015} In fact, detailed thermodynamic studies of the 122 family (Ba$_{1-x}$Na$_x$Fe$_2$As$_2$, Sr$_{1-x}$Na$_x$Fe$_2$As$_2$, BaFe$_2$(As$_{1-x}$P$_x$)$_2$), revealed a multitude of new magnetic phases arising in the vicinity of the superconducting dome, thus providing further insight into the interplay of such phases with superconductivity.\cite{BaNa122_Wang_2016,122P_Boehmer_2012,SrNa122_Wang_2019,BaK122_Boehmer_Nat_comm_2015}
Similar techniques were successfully applied to FeSe crystals, that appear to behave differently from other Fe-based compounds and the effect of nematicity on superconductivity is still debated. As an example, in FeSe$_{1-x}$S$_x$ and Fe$_{1-x}$Co$_x$Se the increase of the superconducting transition temperature ($T_\mathrm{c}$) appears to be followed by an enhanced orthorhombic distortion, in sharp contrast to other Fe-based compounds.\cite{FeSe_review_Boehmer_2017,FeSe_orth_p_Kothapalli_2016,Boehmer_TD_NMR_FeSe_2015}
In this context, a precise determination of the phase diagram of 1111 systems of single crystals is highly desirable, because the understanding of the origin of nematicity and the study of its possible interplay with the magnetic and superconducting phases will give important hints on the nature of the pairing mechanism in this class of high-temperature superconductors.

Despite LaFeAsO being the first reported compound of this class of materials, the lack of macroscopic single crystals has so far hindered a more comprehensive investigation of superconductivity in the 1111 family. The successful growth of macroscopic facetted LaFeAsO single crystals by Solid State Crystal Growth (SSCG) was recently reported.\cite{CoLa1111_SC_growth} The use of single crystals is especially important for this class of materials, for which the study of the anisotropic quantities within the $ab$-plane is crucial to investigate the nematic phase precursing superconductivity.\cite{Xiaochen_ER_CoLa1111,Piotr_NMR_CoLa1111,CoLa1111_Wuttke_Nernst_2020,Caglieris_ER_ES_2020}
In particular, for the purpose of this paper, the improved growth along the $c$ axis with clear facets allows a more precise orientation of the crystals for thermal expansion measurements in order to resolve the small lattice distortion in the $ab$ plane.

In the present work, the phase diagram of Co-doped LaFeAsO was re-investigated on macroscopic faceted single crystals by means of thermodynamic probes. Magnetization and specific heat measurements were used to probe the suppression of the antiferromagnetic SDW transition $T_\mathrm{N}$ as a function of nominal Co content.
The direct substitution of the Fe sites by Co ions, despite inducing a higher degree of structural disorder, with respect to F doping, assures a better control on the electron doping by substituting the dopant directly within the FeAs plane.
In order to investigate the evolution of the orthorhombic distortion and to verify the presence of a long-range nematic phase in the higher-doping region, the temperature dependence of the linear thermal expansion coefficient was obtained in the $a$ and $b$ crystallographic axes for several Co doping compositions. 
We are able to establish important key features within the phase diagram: (i) a gradual suppression of the itinerant antiferromagnetic order up to 5\%~Co content; (ii) the presence of a long-range nematic order transition up to 4.5\%~Co; (iii) no coexistence of magnetism and superconductivity. Finally, using different approaches qualitative statements on the uniaxial pressure dependence of the magnetic transition temperature were obtained.

\section{EXPERIMENTAL DETAILS}

Single crystals of LaFe$_{1-x}$Co$_x$AsO were grown by Solid State Crystal Growth. The growth process resulted in macroscopic single crystals up to 1~x~3~x~0.4~mm$^3$. The growth together with the structural and magnetic characterization of the parent compound have been reported by the authors in a previous work\cite{CoLa1111_SC_growth}, in which the phase purity of the parent compound is demonstrated and discussed in detail. Co doping compositions throughout the series were measured by energy dispersive X-ray diffraction (EDX), giving a good agreement with the nominal compositions. 
Note, that a few preliminary thermal expansion data published previously\cite{CoLa1111_SC_TE} are labelled by the Co contents obtained from the EDX investigations, while nominal compositions will be used in this work and related papers\cite{Xiaochen_ER_CoLa1111,Piotr_NMR_CoLa1111,Hajo_NMR_FLa1111_pc,Piotr_NQR_CoLa1111,CoLa1111_Wuttke_Nernst_2020,Caglieris_ER_ES_2020}. 

The DC magnetization was measured as a function of temperature and magnetic field by means of a superconducting quantum interference device vibrating sample magnetometer (SQUID-VSM) by Quantum Design. For temperature dependent experiments, the sample was cooled down to 1.8 K in zero field and the measurement was performed upon warming from 1.8 to 300 K with an applied external field within the $ab$-plane ($H~\parallel~ab$). In order to probe the superconducting transition, zero-field-cooled (ZFC) and field-cooled (FC) magnetization measurements were performed in an external magnetic field of 10 Oe.

Specific heat ($C_\mathrm{p}$) measurements were performed by the relaxation method using a Physical Property Measurement System (PPMS) from Quantum Design. Prior to each sample measurement, the contribution from the platform of the puck and the grease, used to assure a good thermal contact with the sample, has been measured at 0~T, in order to properly subtract this background from the total heat capacity measured by the PPMS. Note that this method is not able to quantify first-order transitions, where latent heat is involved. However, on a qualitative level structural first-order transitions can still be pinned down.

High-resolution capacitance dilatometry was used to probe the temperature evolution of lattice parameters throughout the Co doped series by orienting the samples along different crystallographic axes. A commercially available cell\cite{Kuechler_2012} was adapted to be used with the PPMS cryostat by Quantum Design in order to sweep temperature (1.8-300~K) and magnetic field (0-9~T). The capacitance signal was detected by a capacitance bridge with a resolution up to 10$^{-5}$~pF. For the present work, the relative length change ($\Delta L/L_0$) was measured for the directions [100]$_\mathrm{T}$ and [110]$_\mathrm{T}$, referring to the high temperature tetragonal phase. The sample was slowly cooled down to 1.8~K and measured in zero field and upon sweeping temperature to 250~K at 0.2~K/min. The background, given by the thermal expansion of the dilatometer cell, is estimated by measuring $\Delta L/L_0$(T) for a Cu reference sample, from which the literature values for pure Cu is subtracted as described in previous reports.\cite{Kuechler_2012}

\begin{figure*}
    \includegraphics[width=0.8\textwidth]{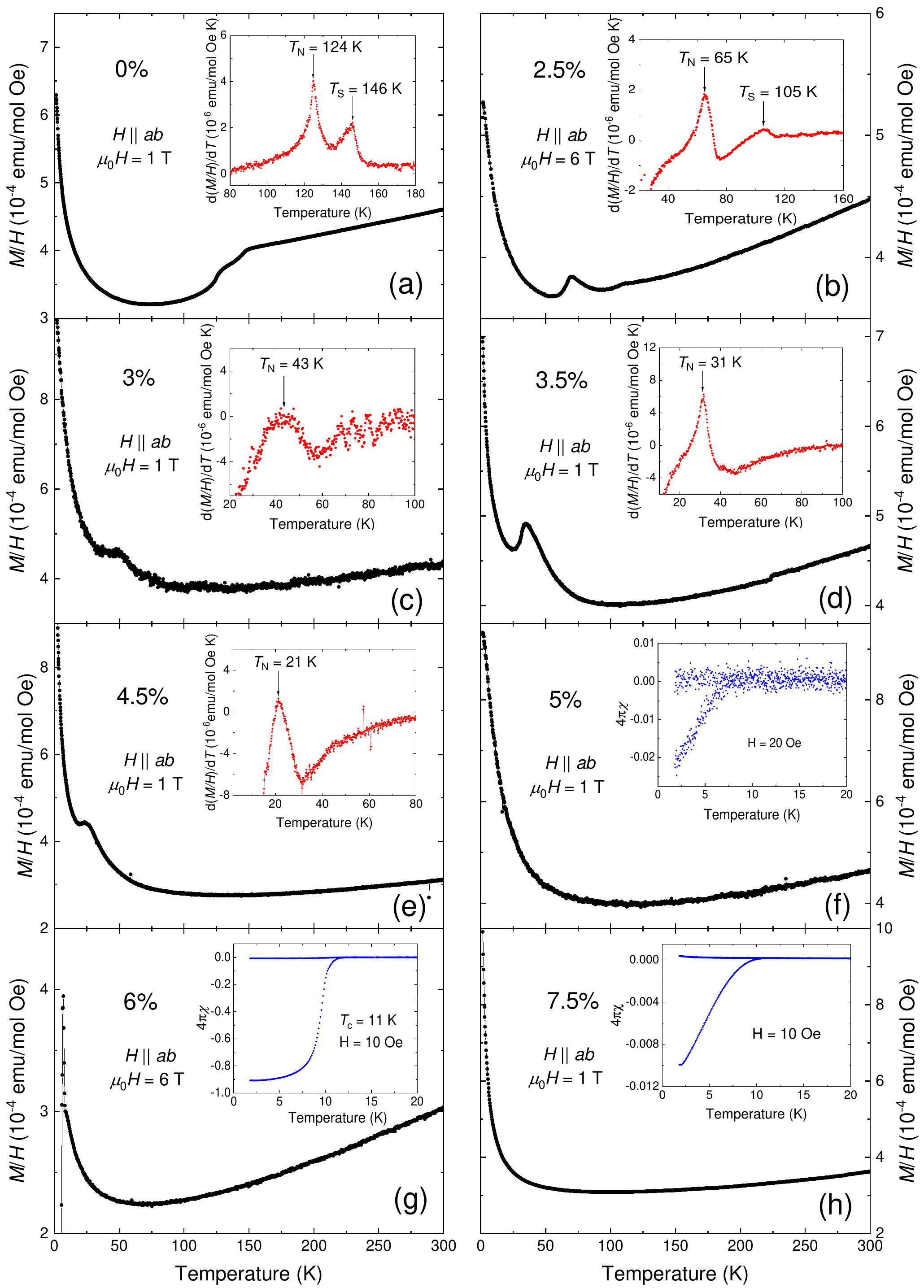}
    \caption{Normalized magnetization $M/H$ for the LaFe$_{1-x}$Co$_x$AsO series: (a) parent compound, (b) 2.5\%, (c) 3\% (Taken from Ref.~\cite{Caglieris_ER_ES_2020}), (d) 3.5\%, (e) 4.5\%, (f) 5\%, (g) 6\%, (h) 7.5\% nominal Co content. A field of $\mu_0H$~=~1~T was applied within the $ab$ plane (in some cases a field of 6~T was applied in order to improve the resolution). The insets for (a)-(e) show the temperature derivative of $M/H$, while for (f)-(h) the volume susceptibility is shown for $H$~=~10~Oe at low temperature for both ZFC and FC measurement conditions\bibnote[Note1]{The higher noise level in 3\% and 5\% Co reflects the lower sample mass of these compositions.}.}
    \label{X_summary}
\end{figure*}

\section{Results}

\subsection{Long-range magnetic order and superconductivity}\label{M}

Fig.~\ref{X_summary} (a)-(h) show the magnetization as function of temperature throughout the Co-doped series from the parent compound to the overdoped 7.5\% Co content. The $\chi(T)$ curve of the parent compound LaFeAsO at $\mu_0H$~=~1~T is shown in Fig.~\ref{X_summary}~(a). At high temperatures, a linear temperature  dependence is visible, well known from the polycrystalline samples and previously interpreted as a manifestation of short-range antiferromagnetic correlations that are still present above the nematic phase.\cite{FLa1111_mag_klingeler} As the temperature is lowered two clear anomalies are visible in the magnetization curve below 200~K. The use of single crystals allows for an improved resolution in the measurements, while much broader transitions were observed in previous reports, as expected from the use of polycrystalline samples \cite{FLa1111_mag_klingeler,CoLa1111_PC_Sefat_2008,CoLa1111_PC_2009}. The inset of Fig.~\ref{X_summary}~(a) depicts the derivative $\frac{dM/H}{dT}$, where two sharp peaks are visible at 146~K and 124~K, assigned to the structural and magnetic transitions, respectively.

Upon substituting Fe by Co atoms in the FeAs planes, the two distinct anomalies in $\chi(T)$ persist up to 2.5\%~Co content (Fig.~\ref{X_summary}~(b)), where both transition temperatures appear to be significantly lowered. It has to be noted that, while the low-temperature magnetic anomaly remains relatively sharp and clearly visible in the magnetization curve, the high-temperature structural transition appears to be progressively broadened with increasing Co content. In fact, for 0.025~$<$~$x$~$<$~0.05 Co only one single anomaly can be discerned. Following the aforementioned observation as well as comparing with thermal expansion measurements (see Section~\ref{DLL}), the single peak anomalies found in the intermediate doping region have been assigned to the SDW ordering (Fig.~\ref{X_summary}~(c)-(e)). This magnetic transition temperature shows a strong and continuous lowering from $\sim$~124~K in the parent compound to $\sim$~20~K in the 4.5\% Co doped samples, followed by a complete suppression of the anomaly for the 5\% and higher Co-doped samples. These results are in good agreement with measurements of the relaxation rate (1/$T_1T$) by nuclear magnetic resonance (NMR) spectroscopy performed on the same batch of samples\cite{Piotr_NMR_CoLa1111}. Notably, although $T_\mathrm{N}$ is gradually suppressed, the magnetic anomaly in $\chi(T)$ does not seem to significantly broaden with Co doping. This behavior is in agreement with the results reported by Prando \textit{et al.}\cite{Prando_Co_dilution} for CeFeAsO, where the doping dependence of the magnetic anomaly is not changed with respect to in-plane (Fe$_{1-x}$Co$_x$) or out-of-plane (O$_{1-x}$F$_x$) substitution and the phase boundaries sit on top of each other. This suggests that the SDW instability in the 1111 family is surprisingly stable with respect to in-plane disorder.  In fact, in our $\chi(T)$ curves, clear magnetic anomalies can be probed until their complete suppression at $\sim$~5\% Co doping, while the structural transition is continuously broadened with increasing Co content.

\begin{figure}[t]
	\center
	\includegraphics[width=0.45\textwidth]{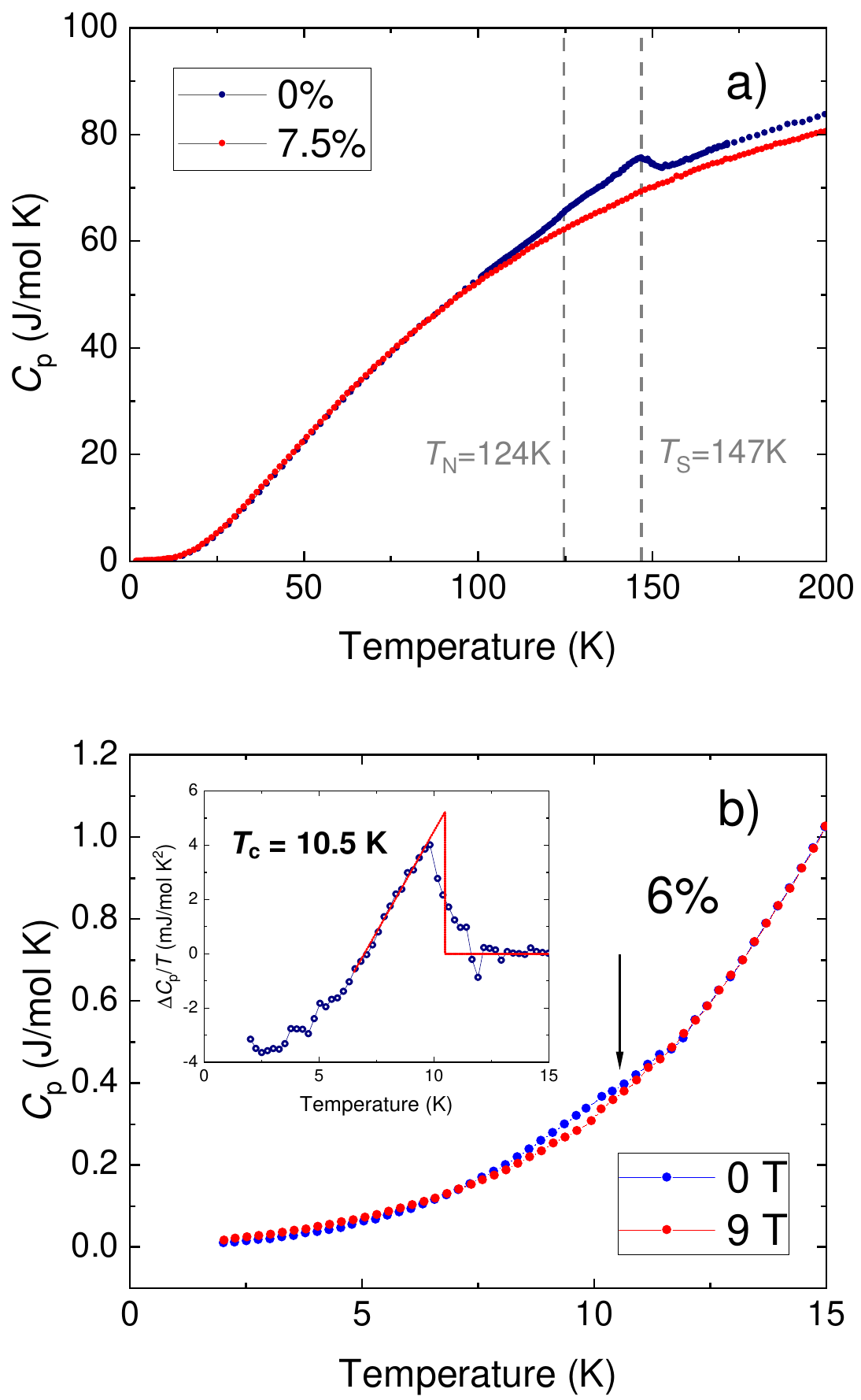}
	\caption[width=0.45\textwidth]{Specific heat of LaFe$_{1-x}$Co$_x$AsO for (a) the parent compound and for 7.5\% Co as well as (b) for 6\% Co doping. The inset shows the determination of the superconducting transition temperature from the subtracted signal $\Delta C_\mathrm{p}/T~=~\frac{C_\mathrm{p(0T)}-C_\mathrm{p(9T)}}{T}$, approximating the specific heat contribution from the superconducting phase. The offset in $C_\mathrm{p}$ for $T~>~150~K$ between the 0\% and 7.5\% Co is most probably due to the difference in the sample coupling at high temperature.}
	\label{Cp_summary}
\end{figure}

The insets of Fig.~\ref{X_summary}~(f)-(h) show the superconducting transition for 5, 6 and 7.5\% Co compositions. The transition temperatures were determined by a linear construction close to the bifurcation point between the ZFC and FC curves. Bulk superconductivity can be observed for the 6\% Co sample (inset of Fig.~\ref{X_summary}~(g)), with the highest transition temperature $T_\mathrm{c}~\sim$~11~K. This is in good agreement with the values previously found for polycrystalline samples.\cite{CoLa1111_PC_2009,CoLa1111_PC_Sefat_2008} For the neighboring compositions a volume fraction $\lesssim$~2\%  suggests spurious superconductivity, possibly related to surface superconductivity or too small superconducting regions rising from an inhomogeneous doping distribution within the sample. The latter observation would suggest that superconductivity in Co-doped La1111 develops in a very confined doping region around 6\%~Co content. This scenario is supported by means of resistivity measurements as a function of temperature reported by Hong \textit{et al.}\cite{Xiaochen_ER_CoLa1111}, suggesting a fast suppression of the superconducting phase in the immediate vicinity of the 6\% composition. Also, NMR spectroscopy measurements show evidence for short-range magnetic ordering in 4.2\% and 5.6\% Co samples, which would further support the presence of inhomogenous behavior in this doping region.\cite{Piotr_NMR_CoLa1111}
It has to be noted that, while electron doping is necessary to induce superconductivity, in-plane disorder introduced by Co doping proved to be detrimental for this phase in 1111 systems with respect to out-of-plane F doping, as also demonstrated for CeFeAsO\cite{Prando_Co_dilution}, showing significantly lowered critical temperatures.

Fig.~\ref{Cp_summary} shows the heat capacity of some representative compositions. The total heat capacity is composed of
\begin{equation}
C_p=C_{el}+C_{ph}+C_{mag},
\label{eq:OP}
\end{equation}
representing the electronic ($C_{el}$), phononic ($C_{ph}$) and magnetic ($C_{mag}$) contributions.

Fig.~\ref{Cp_summary} (a) shows the heat capacity results for the parent compound and for a 7.5\% Co-doped sample. The undoped LaFeAsO sample, reported in our previous work\cite{CoLa1111_SC_growth}, has a clear additional contribution above the background ($C_\mathrm{el}+C_\mathrm{ph}$) for T~$>$~100~K. Additional $C_p$ contributions are mainly observed for the structural phase transition at $T_\mathrm{S}$~=~147~K, while a smaller anomaly at lower temperature can be assigned to the ordering of the SDW phase at $T_\mathrm{N}$~=~124~K.\cite{CoLa1111_SC_growth} In contrast, for the highest doped compound (Fig.~\ref{Cp_summary}~(a)), no additional entropy contributions can be recorded in $C_\mathrm{p}$, signaling the suppression of the magnetic and nematic phases for 7.5\%~Co substitution, in agreement with our $\chi(T)$ measurements.

The low-temperature specific heat for the 6\%~Co-doped sample is shown in Fig.~\ref{Cp_summary}~(b) for zero field and for an external field of 9~T applied along the $c$ axis. In order to extract the SC transition temperature, the 9~T curve was used as an approximation for the normal-state specific heat capacity for $T~>$~2~K, yielding a superconducting transition temperature $T_\mathrm{c}$~=~10.5~K via an entropy-conserving linear construction (see the inset of Fig.~\ref{Cp_summary}~(b)). It has to be noted that an external magnetic field of 9~T shifts the superconducting transition temperature to much lower temperature $T~<$~2~K, such that the electronic specific heat contribution from the superconducting phase can be neglected in the vicinity of the 0~T transition around 10~K, while the electronic and phononic terms should remain approximately unchanged by the field. The manifestation of an anomaly at $T_\mathrm{c}$ in the $C_\mathrm{p}$ studies of this composition further supports the bulk character of superconductivity. Correspondingly, the lack of a detectable anomaly in $C_\mathrm{p}$ for the 7.5\% Co-doped sample is consistent with the small volume fraction of SC observed via $\chi$.

\subsection{Orthorhombic lattice distortion}\label{DLL}

\begin{figure}
	\center
	\includegraphics[width=0.55\textwidth]{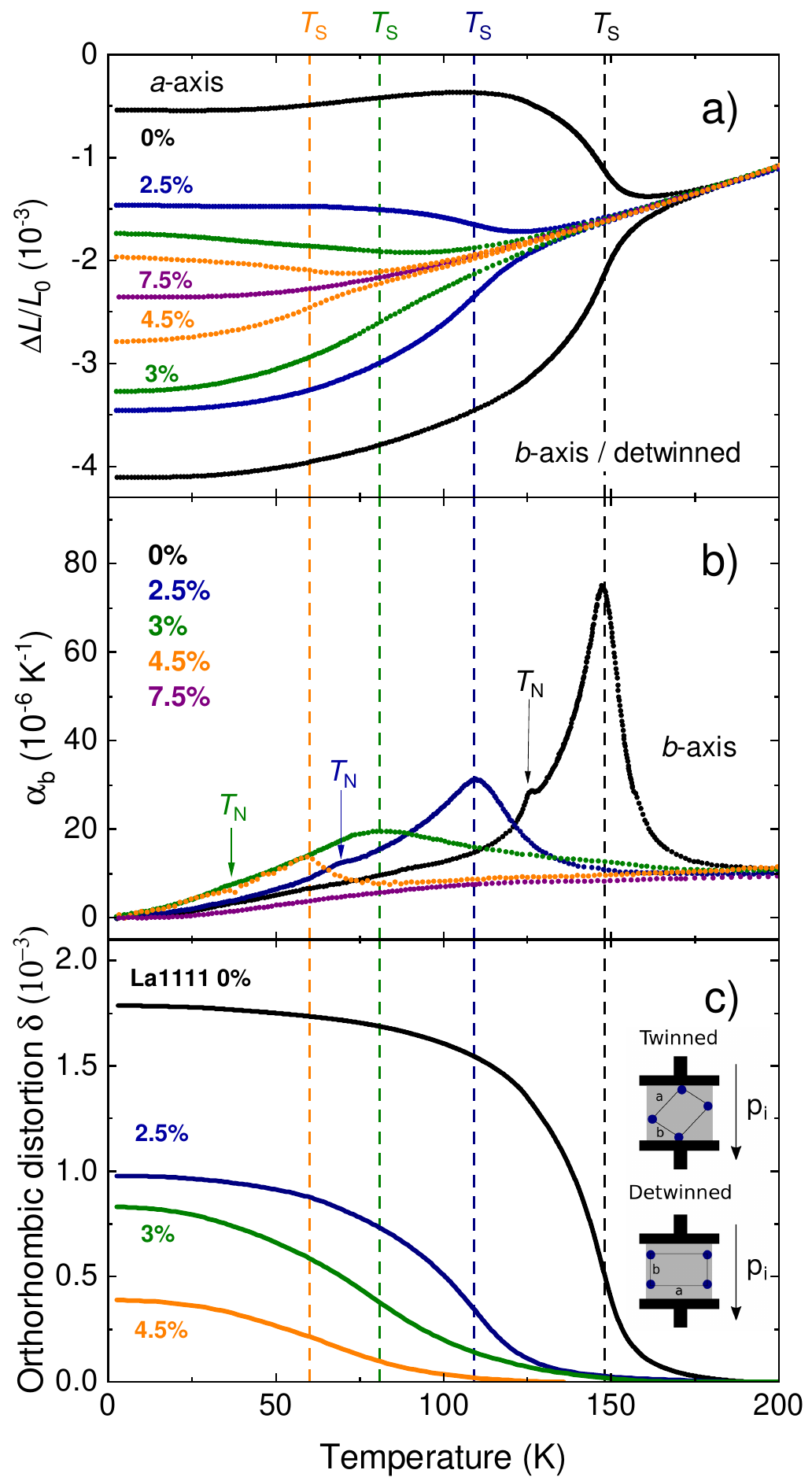}
	\caption{Thermal expansion of the LaFe$_{1-x}$Co$_x$AsO series: (a) normalized lengths changes $\Delta L/L_0$, (b) linear thermal expansion coefficient $\alpha_b$, and (c) orthorhombic order parameter $\delta$. The dashed lines represent the structural transition temperatures $T_\mathrm{S}$ for different Co contents, while the magnetic transition temperatures $T_\mathrm{N}$ are marked by arrows in (b). The inset of (c) shows a schematic representation of the twinned and detwinned measurement configurations. The curves for 0, 2.5, 3\% Co are taken from Ref.~\cite{CoLa1111_SC_TE}. Note that the pressure applied to each crystal in the series varies due to the specific geometry and dimensions of each sample.}
	\label{TE_summary}
\end{figure}

In order to probe the orthorhombic lattice distortion, we performed thermal expansion measurements on some representative compositions throughout the series, thereby extracting the relative lengths changes $\Delta L/L_{0}$ as a function of temperature, where $L_0$ is the sample length at 300K. The linear thermal expansion coefficient can be obtained as the derivative of the lengths changes as
\begin{equation}
\alpha_i = \frac{1}{L_{0}} \left( \frac{\partial L_i}{\partial T}\right)_{p_i},
\label{eq:OP}
\end{equation}
where $i$ represents the $a$ or $b$ crystallographic axis in the orthorhombic phase in our work.

Due to the particular design of the dilatometer, a small uniaxial force is always applied to the sample in order to fix it to the capacitor plates and it can be used to actively detwin the crystals, as demonstrated by previous reports.\cite{122P_Boehmer_2012,BaK122_Boehmer_Nat_comm_2015,BaNa122_Wang_2016} By orienting the crystal in the [110]$_T$ direction in the tetragonal phase, a so-called detwinned measurement is performed yielding $\Delta b/b_0$ in the orthorhombic unit cell, while the application of a small force in the [100]$_T$ direction leads to a signal (twinned measurement) corresponding to an average of the two axes, i.e., $\frac{1}{2} \left(\frac{\Delta a}{a_0} + \frac{\Delta b}{b_0} \right)$. The lengths changes of the $a$ parameter ($\Delta a/a_0$) are obtained by the subtraction of the two signals (schematics of used sample orientations are shown in the inset of Fig.~\ref{TE_summary}~(c)). Fig.~\ref{TE_summary}~(a) depicts $\Delta L/L_0$ for the two crystallographic main directions in the $ab$ plane. For the parent compound LaFeAsO a clear splitting between the $a$ and the $b$ axis can be observed for $T~\le$~180~K and it is maintained down to low temperature, signaling the nematic phase. The lattice distortion is progressively reduced and substantially broadened in amplitude by electron doping, but is still present up to 4.5~\% Co content. The precise determination of the transition temperatures can be obtained from the thermal expansion coefficient $\alpha$ (shown in Fig.~\ref{TE_summary} (b)). For the parent compound one clear anomaly can be detected at $T_\mathrm{S}$~=~147~K, corresponding to the structural transition, followed by a smaller peak indicating long-range magnetic ordering at $T_\mathrm{N}$~=~124~K. These anomalies are strongly suppressed and broadened upon Co doping, with $T_\mathrm{S}$ shifting from $\sim$~147~K down to $\sim$~60~K for samples with 4.5~\% Co content. In magnetization measurements (see sect.~\ref{M}) the structural phase transition could not be detected for $x~>$~2.5~\% Co doping, pointing at a weak coupling between lattice and spin degrees of freedom together with an intrinsic broadening in temperature upon in-plane dilution with Co. In contrast, for the highly-doped 7.5~\% Co sample, twinned and detwinned signals lie on top of each other, indicating that C$_4$ symmetry is maintained in the full temperature range and that the nematic phase is completely suppressed.

Fig.~\ref{TE_summary}~(c) presents the extracted orthorhombic distortion order parameter. By knowing the relative lengths changes and the structural parameter $a_\mathrm{0}$ in the tetragonal phase determined by XRD diffractometry at room temperature, the temperature evolution of the lattice parameters can be calculated. The orthorhombic order parameter is then given by
\begin{equation}
\delta=\frac{a-b}{a+b},
\label{eq:OP}
\end{equation}
with $a$ and $b$ referring to the in-plane lattice parameters in the orthorhombic phase.
The magnitude of the distortion is progressively lowered with increasing Co content. As reported by Wang \textit{et al.}\cite{CoLa1111_SC_TE}, the overall distortion of 1.8~x~10$^{-3}$ (in the limit $T~\to~0$) for the parent compound is considerably smaller than the values found for other IBS. By comparing with previous works, we noted that the $\delta$ for LaFeAsO is reduced by a factor of 0.5 and 0.6 is found compared to BaFe$_2$As$_2$\cite{BaK122_orth_neutron_Avci_2011,BaNa122_Wang_2016,122P_Boehmer_2012} and FeSe\cite{FeSe_review_Boehmer_2017,FeSe_orth_p_Kothapalli_2016,Boehmer_TD_NMR_FeSe_2015}, respectively.

Also, for all measured compositions with $x$~=~0~-~4.5\%~Co the onset of the orthorhombic distortion happens at higher temperatures with respect to the determined $T_\mathrm{S}$ (as shown in Fig.~\ref{TE_summary}~(a), (c)).
In fact, for the parent compound, as shown in Fig.~\ref{TE_summary}~(b), the peak associated with the structural transition in $\alpha_\mathrm{b}$ signals $T_\mathrm{S}$, while the transition width extends for $\sim$~10~K above $T_\mathrm{S}$. Such a phenomenon was also seen in polycrystalline F doped LaFeAsO samples\cite{FLa1111_TE_Liran_2009,Hess_FLa1111_PC}, interpreted as the softening of the lattice in the $ab$ plane due to a precursor state of nematic fluctuations above the long-range nematic phase.
In our case though, different contributions might take part in the overall broadening in temperature of the structural transition: (i) the presence of fluctuations preceeding the long-range transition, allowing a partial distortion of the tetragonal symmetry already above the transition temperature; (ii) the small uniaxial pressure applied for the thermal expansion measurement along the $b$ axis used to detwin the sample\bibnote[Note2]{Due to the sample size and geometry, the applied pressure can also vary between different compositions upon measuring.}; (iii) ultimately, Co/Fe substitution inducing disorder in the $ab$ plane. In order to disentangle these effects, the thermal expansion coefficient was measured as a function of temperature at different applied uniaxial pressures.

\begin{figure}
	\center
	\includegraphics[width=0.49\textwidth]{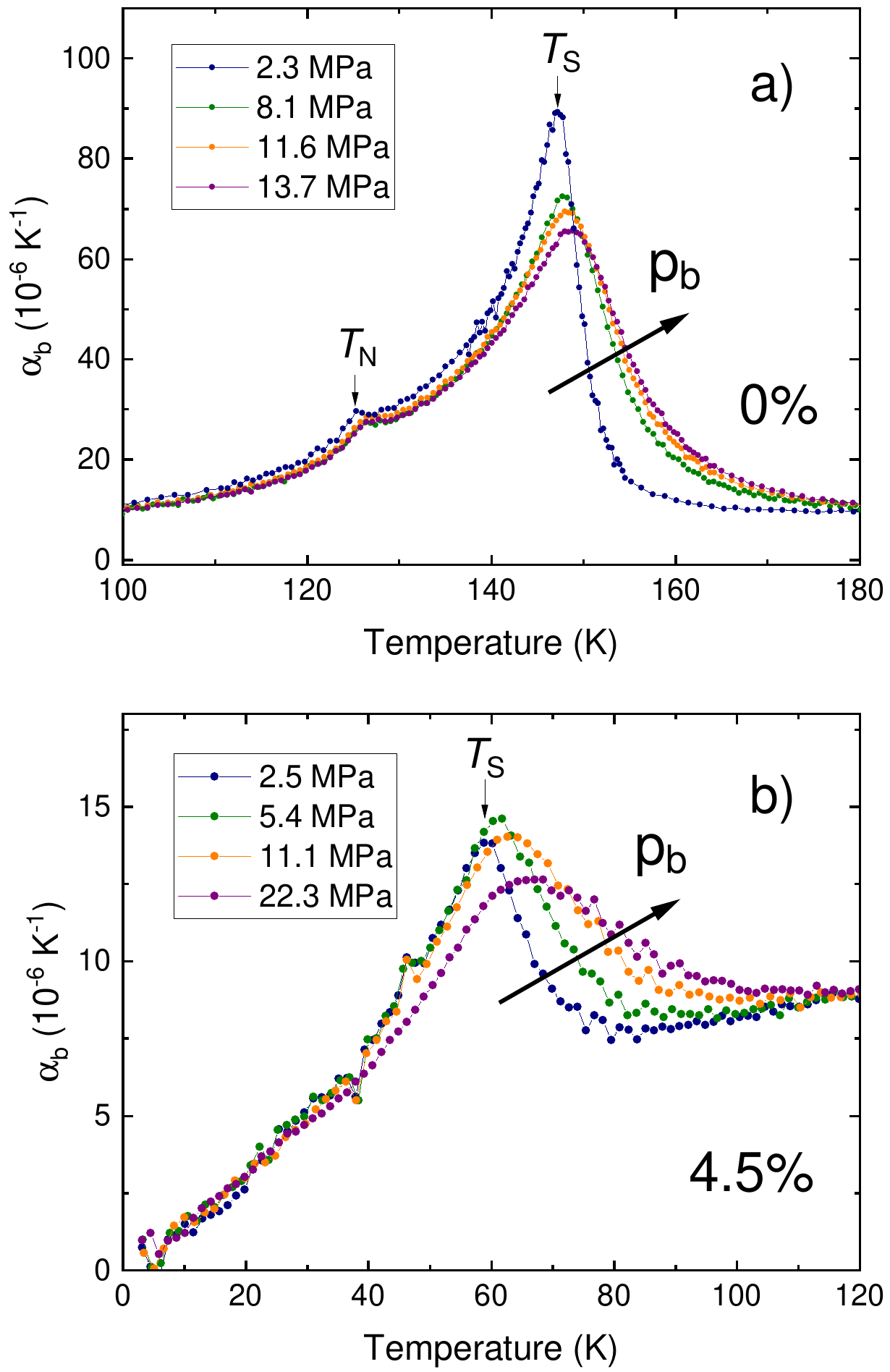}
	\caption{Uniaxial pressure dependence of the thermal expansion coefficient in the detwinned direction ($\alpha_b$) of LaFe$_{1-x}$Co$_x$AsO for (a) the parent compound and (b) 4.5\% Co doping\bibnote[Note3]{For 4.5\% Co, the peak at the structure transition upon application of 2.5 MPa (Fig.~\ref{p_dependence_summary}(b)) shows an artificially decreased amplitude compared to the higher pressure 5.4~MPa due to a small tilting of the sample towards the $c$ axis, resulting from the very small uniaxial force applied. Nevertheless, for the purpose of our qualitative analysis, the curve still clearly shows distinctive features that can be correctly discussed in terms of position and width of the peak.}. It has to be noted that the application of increasing uniaxial pressure does not substantially increase the overall orthorhombic distortion, confirming the almost complete detwinning of the measured samples.}
	\label{p_dependence_summary}
\end{figure}

\subsection{Pressure dependence of $T_\mathrm{S}$ and $T_\mathrm{N}$}\label{p_dependence}

Fig.~\ref{p_dependence_summary} shows the uniaxial pressure dependence of the thermal expansion curves in the detwinned direction ($\Delta b/b_0$) for 0 and 4.5\%~Co doping. The structural anomaly in LaFeAsO is progressively broadened and shifted to higher temperature, as expected from the application of uniaxial pressure along the $b$ axis.\cite{Ba122_P_dep_Dhital_2012,Ba122_P_dep_Lu_2016} 
Notably, the same behavior can be seen for the 4.5\% Co sample. In particular, by decreasing the applied uniaxial pressure ($p_\mathrm{b}$) down to $\sim$~2.5~MPa, the peak connencted to the structural distortion becomes sharper, eventually resembling a $\lambda$-shaped anomaly indicative for a second-order phase transition, thus confirming the long-range character of the nematic transition in the range of 0\% to 4.5\% Co doping. Interestingly, in the parent compound (Fig.~\ref{p_dependence_summary}(a)), for which the broadening effects due to Co substitution can be neglected, the high-$T$ tail is still visible at the lowest applied pressure (2.3~MPa). Nevertheless, it appears considerably reduced, hindering clear-cut conclusions about the intrinsic effect of a precursor fluctuation regime above $T_\mathrm{S}$.

\begin{figure}[t]
	\center
	\includegraphics[width=0.55\textwidth]{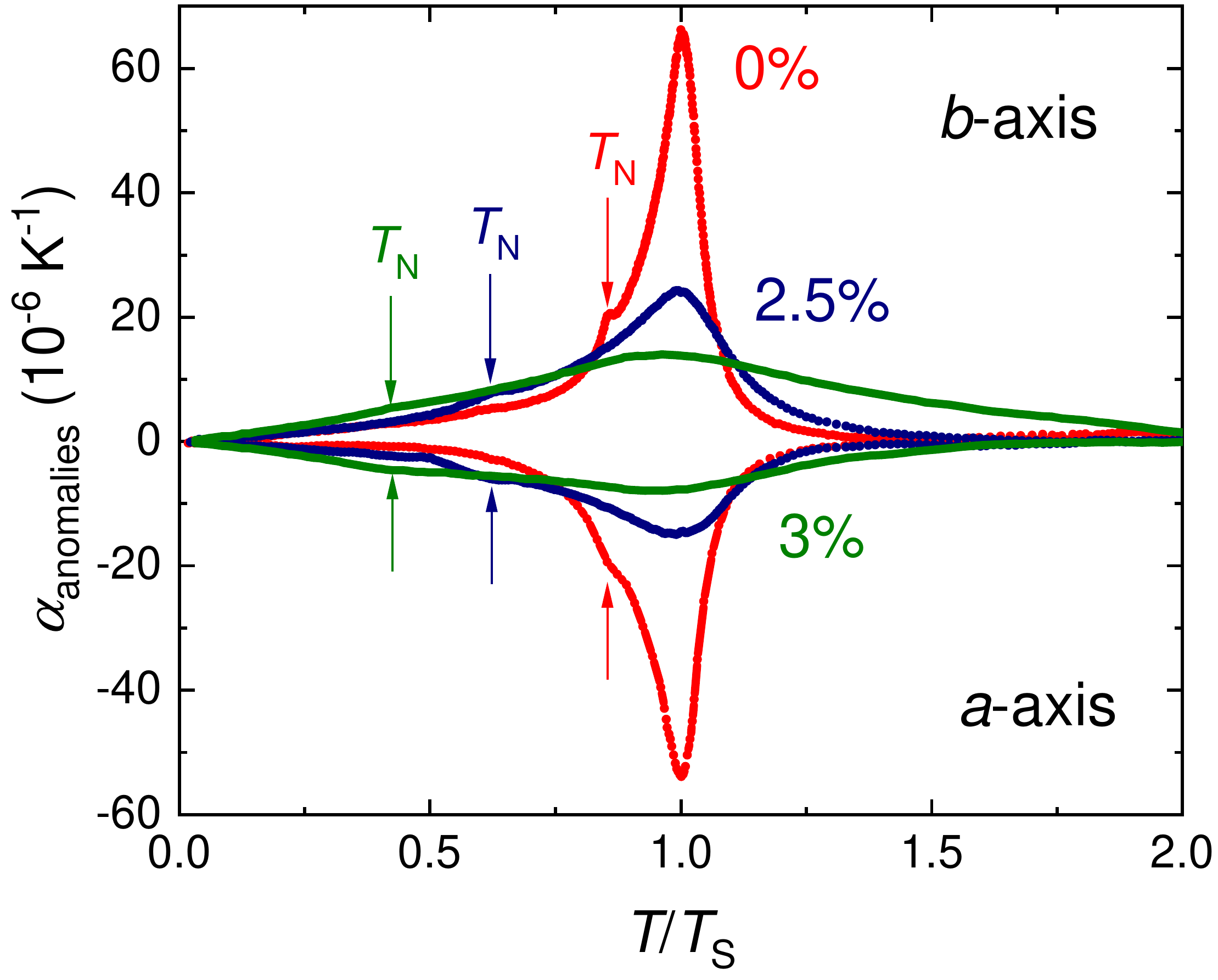}
	\caption{Anomalies in the linear thermal expansion coefficient for the $a$ and $b$ axes for 0, 2.5 and 3\%~Co content. The non-magnetic contribution was approximated by using the thermal expansion coefficient obtained for the overdoped 7.5\%~Co, which does not show any transition in any of the thermodynamic quantities under studies.}
	\label{TE_anom_Tn}
\end{figure}

The uniaxial pressure dependence of second-order phase transitions, in the case of jump-like anomalies, can be extracted by combining specific heat and thermal expansion measurements through the Ehrenfest relation,
\begin{equation}
\frac{dT_\mathrm{N}}{dp_i} = T_\mathrm{N} V_m \frac{\Delta\alpha_i}{\Delta C_p},
\label{eq:OP}
\end{equation}
where $\Delta\alpha_i$ and $\Delta C_\mathrm{p}$ correspond to the height of the thermal expansion and the specific heat anomaly, respectively ($i$ denotes the crystallographic direction), $V_m$ is the molar volume, and $T_\mathrm{N}$ is the transition temperature into the SDW state. Fig.~\ref{TE_anom_Tn} shows a summary of the thermal expansion anomalies for the $a$ and $b$ axes up to 3\% Co content (note that the curve for the 4.5\% Co-doped sample was excluded because the magnetic anomaly found in $\chi(T)$ could not be resolved in $\alpha$ due to the small amplitude of the related lattice distortion).
The non-magnetic contributions (approximated by the thermal expansion curve of the 7.5\% Co doped sample) were subtracted in order to obtain $\alpha_\mathrm{anomalies}$.
In our measurements the entropy changes at the magnetic ordering temperature appears to be much smaller than at the structural transition in both $C_\mathrm{p}$ and $\alpha_i$, making the determination of $\Delta C_\mathrm{p}$ and $\Delta\alpha_i$ on a quantitative level difficult and subject to big uncertainties. Still, from the sign of the anomalies ($\Delta C_\mathrm{p}~>~0$, $\Delta\alpha_a~<~0$, $\Delta\alpha_b~>~0$), an opposite effect can be expected for the transition temperature for uniaxial pressure in the $a$ and $b$ directions, i.e., $\frac{dT_\mathrm{N}}{dp_\mathrm{a}}~<~0$ and $\frac{dT_\mathrm{N}}{dp_\mathrm{b}}~>~0$ (see Fig.~\ref{TE_anom_Tn}). Furthermore, the similar size of the magnetic anomalies in the phonon subtracted thermal expansion coefficient ($\alpha_\mathrm{anomalies}$) suggests that the in-plane pressure dependence is most probably comparable in size for the $a$ and $b$ axis. Interestingly, the volume thermal expansion was previously measured on LaFeAsO polycrystals by Wang \textit{et al.}\cite{FLa1111_TE_Liran_2009}, where an overall negative hydrostatic pressure dependence was determined for $T_\mathrm{N}$.

\section{Phase diagram}\label{La1111_PD}

\begin{figure}[t]
\centering
\begin{minipage}[c]{\columnwidth}
\centering
    \includegraphics[width=0.6\linewidth]{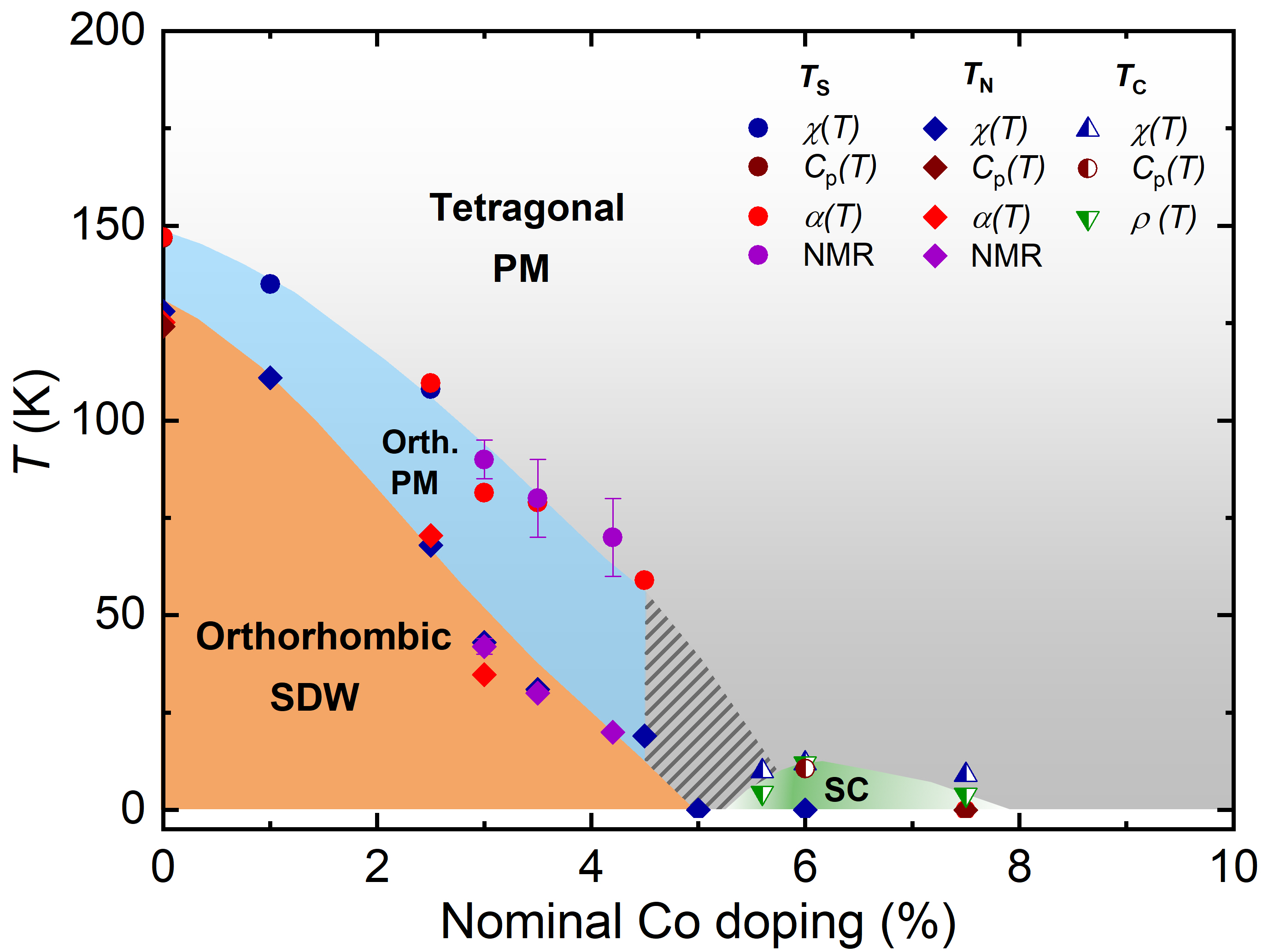}
    \caption{Phase diagram of Co-doped LaFeAsO single crystals. Points from NMR spectroscopy and resistivity on the same series of single crystals were added from Refs.~\cite{Piotr_NMR_CoLa1111} and~\cite{Xiaochen_ER_CoLa1111}, respectively. The shaded area within the orthorhombic/paramagnetic phase represents the doping region (4.5~-~6~\%~Co) where direct lattice probes are not available. The phase boundaries for the nematic state in this region are extrapolated from the data points of neighboring compositions. The color code in the superconducting dome reflects the rapid decrease of volume fraction estimated from magnetization measurements around the optimally doped 6\% Co content, as described in Sect. \ref{M}.}
    \label{La1111_PD_sc}
\end{minipage}
\end{figure}

Fig.~\ref{La1111_PD_sc} shows the $x-T$ phase diagram of LaFe$_{1-x}$Co$_x$AsO, which summarizes the transition temperatures extracted from different thermodynamic probes on our substitution series. The system proves to be very sensitive to Co doping and the transition temperatures $T_\mathrm{S}$ and $T_\mathrm{N}$ are continuously suppressed by $\sim$~100~K up to 4.5\% Co doping, while both phases are not present anymore in the 7.5\% Co-doped samples. The suppression of the magnetic SDW phase can be inferred from the disappearance of the respective anomaly in the $\chi(T)$ curves for $x~\geq$~5\% (Fig.~\ref{X_summary}~(f)-(h)), suggesting the full suppression of long-range magnetic order between 4.5\% and 5\% doping. Superconductivity was found in a small region of Co doping around 6\%~Co with a maximum $T_\mathrm{c}$~=~10.5~K.

Different from our results, Wang \textit{et al.}\cite{CoLa1111_PC_2009} found that in polycrystalline Co-doped La1111 the orthorhombic/SDW phase could be detected only between 0\% and 2.5\% Co and superconductivity develops as a separate phase for $x~\geq$~2.5\%.
A similar behaviour was found in previous reports\cite{FLa1111_Luetkens,Hess_FLa1111_PC} on F-doped polycrystalline samples, with a modest suppression of the long-range nematic order at (nominal) low-doping concentrations, which is lost abruptly at $\sim$~5\% F content in favor of a short-range nematic order and a broad superconducting dome.
The apparent discrepancy with respect to our data can possibly be attributed to an increased resolution obtained by measuring single crystals directly in the $ab$ plane, thus assessing more subtle features and anisotropies with respect to magnetic and structural changes.
Also, more recent works\cite{Lang_NQR_RE1111_2016,Hajo_NMR_FLa1111_pc} reported a generalized phase diagram for F-doped $RE$FeAsO polycrystals, where the actual doping content in the FeAs plane is estimated from the spectral weight transfer between different peaks in the nuclear quadrupole resonance (NQR) spectra. Through a careful correction of the doping content it was shown, that the SDW phase in La-based systems is gradually suppressed with increasing doping, with no reported coexistence between long-range magnetism and superconductivity, in good agreement with the data reported in this work.

The suppression of the magnetic SDW phase by $\sim$~5\% Co is also in agreement with a recent publication by Hong \textit{et al.}\cite{Xiaochen_ER_CoLa1111} about elastoresistance (ER) measurements on the same batch of Co-doped single crystals. The curves show a local maximum in the nematic susceptibility ($\chi_\mathrm{nem}$) around 4\% Co content, which may be explained by the suppression of the long-range magnetic order around that doping value, preceded by an increase of spin fluctuations.

A nuclear quadrupole resonance analysis of F-doped La1111 showed that for La1111 the electron content is not increased in a uniform way within the FeAs planes upon F doping, but instead intrinsic electronic phase separation occurs on the nanoscale, resulting in high-doping-like (HD) and low-doping-like (LD) regions.\cite{La1111_NQR_Lang_2010,Lang_NQR_RE1111_2016} The same behavior was recently found for Co-doped La1111 polycrystals.\cite{Piotr_NQR_CoLa1111}
It has been shown that the AFM/nematic order develops in the LD-like regions of the sample, while it is not present in the HD-like ones, which are in turn most probably hosting superconductivity.\cite{Hajo_NMR_FLa1111_pc,Piotr_NQR_CoLa1111} Thus, the gradual suppression of the SDW phase and the onset of superconductivity was interpreted as related to a percolation threshold for the HD-like regions. From our pressure dependent studies (see sect.~\ref{p_dependence}), the sharpening of the structural transition for decreasing pressure in both the parent compound and the 4.5\% samples strongly suggest the persistence of the long-range nematic order up to at least 4.5\% Co, but it is fully suppressed at 7.5\%~Co doping.
Therefore, it is still unclear if the long-range structural order can be sustained for the compositions in the vicinity of the superconducting dome around 6\% Co, and whether such lattice distortion would still be present in the non-magnetic state.
On basis of our results and of literature, it seems probable that the long-range character of the nematic order would be lost above a certain Co doping composition, corresponding to the spatial growth of the HD-like regions for x~$\geq$~5\% Co (shaded area in Fig.~\ref{La1111_PD_sc}), as would be suggested by recent NMR measurements for x=0.056.\cite{Piotr_NMR_CoLa1111}. However, further experimental proof is needed to clarify this point.

Differences and similarities can be found with other families of IBS: (i) the substantial phase separation between long-range magnetism and superconductivity is in sharp contrast to a rather broad coexistence region found for 122 compounds, e.g. Ba(Fe$_{1-x}$Co$_x$)$_2$As$_2$\cite{Ba122_coexistence_Wiesenmayer_2011,FePn_Review_Stewart_2011,Co122_XRD_Nandi_2010}; (ii) on the other hand, although clear thermal expansion measurements in proximity to the optimally doped 6\% Co sample are still missing, the phase boundaries derived for the neighbouring compositions suggest an overall competing interaction between long-range nematic order and superconductivity, as suggested by Hong \textit{et al.}\cite{Xiaochen_ER_CoLa1111}, in contrast to what was found for S doped FeSe. 
We conclude that, La1111 seems to present distinctive characteristics that are different in other Fe-based superconductors.

For future works, it will be interesting to investigate the behavior of the nematic phase in close proximity to the superconducting dome in single crystals of Co-doped La1111, in order to further clarify the interplay or competition between these phases, as well as to compare these results with other doping variants on single crystals, e.g., out-of-plane F doping in La1111. 
Such a comparison could also lead to a better understanding of the general pairing mechanism of superconductivity in iron-based compounds.

\section{Summary}

In this work, we report the revisited phase diagram of macroscopic faceted Co-doped LaFeAsO single crystals by thermodynamic probes. The phase diagram proves to be substantially different from other families of Fe-based superconductors. In fact, for the parent compound, clear magnetic and structural transitions can be distinguished. The long-range spin density wave phase was found to be completely suppressed for 5\% Co content, while superconductivity develops in a narrow doping region around 6\% Co doping, thus showing no evident coexistence between the two phases, contrary to what was observed for 122 systems. In addition, the use of single crystals allowed us to assess the small orthorhombic distortion caused by the nematic ordering by means of high-resolution capacitance dilatometry. These studies reveal a continuous reduction of the structural transition temperature $T_\mathrm{S}$ upon doping up to 4.5\% Co and a complete suppression for 7.5\% Co doping, in contrast to previous reports on F/Co doped La1111 polycrystalline samples.  

\section{Acknowledgements}

The authors acknowledge S. Gass, C. G. F. Blum and S. M\"uller-Litvanyi (IFW Dresden) for technical support, Hans-Henning Klauss, Anna E. B\"ohmer and Christoph Meingast for fruitful scientific discussions. This work has been supported by the Deutsche Forschungsgemeinschaft (DFG) through the Research training group DFG-GRK 1621, through SFB 1143 (project id 247310070), through project KL1824/6-1 and the W\"urzburg-Dresden Cluster of Excellence on Complexity and Topology in Quantum Matter -- \textit{ct.qmat} (EXC 2147, project id 390858490).This project has received funding from the European Research Council (ERC) under the European Union’s Horizon 2020 research and innovation programme (Grant Agreement No. 647276-MARS-ERC-2014-CoG). S. Aswartham acknowledges financial support from the Deutsche Forschungsgemeinschaft (DFG) under Grant No. AS 523/4-1 and L. Wang under Grant No.WA4313/1-2.

\bibliography{Ref_thesis2}

\end{document}